\documentclass[twocolumn,showpacs,pre,aps,superscriptaddress]{revtex4}
\usepackage[latin1]{inputenc}
\usepackage{bm}
\usepackage{multirow}
\usepackage{amssymb}
\usepackage{amsbsy}
\usepackage{amsmath}
\usepackage{stmaryrd}
\usepackage{graphicx}
\usepackage{epsfig}

\usepackage{subfigure}

\begin{document}

\title{Finite size effects on transport coefficients for models of atomic wires coupled to phonons}

\author{Christian Bartsch}

\email{cbartsch@uos.de}

\affiliation{Fachbereich Physik, Universit\"at Osnabr\"uck,
             Barbarastrasse 7, D-49069 Osnabr\"uck, Germany}

\author{Jochen Gemmer}

\email{jgemmer@uos.de}

\affiliation{Fachbereich Physik, Universit\"at Osnabr\"uck,
             Barbarastrasse 7, D-49069 Osnabr\"uck, Germany}

\date{\today}

\begin{abstract}

We consider models of quasi-$1$-d, planar atomic wires consisting of several, laterally coupled rows of atoms, with mutually non-interacting electrons. This electronic wire system is coupled to phonons, corresponding, e.g., to some substrate. We aim at computing diffusion coefficients in dependence on the wire widths and the lateral coupling. To this end we firstly construct a numerically manageable linear collision term for the dynamics of the electronic occupation numbers by following a certain projection operator approach. By means of this collision term we set up a linear Boltzmann equation. A formula for extracting diffusion coefficients from such Boltzmann equations is given. We find in the regime of a few atomic rows and intermediate lateral coupling a significant and non-trivial dependence of the diffusion coefficient on both, the width and the lateral coupling. These results, in principle, suggest the possible applicability of such atomic wires as electronic devices,
  such as, e.g., switches. 

\end{abstract}


\maketitle

%
%

\section{Introduction} \label{sec-intro}

The strong desire for an ongoing miniaturization of electronic circuits
has led to the idea of molecular electronics. Very roughly molecular
electronics include all implementations of basic electronic functional
devices on the scale of atoms or molecules. The most basic device surely
is a wire, meaning a system in which an electronic current may flow as a
response to an electric field in a prescribed direction. Various
implementations of such wires have been suggested and experimentally
explored.

Some suggestions are based on chemically synthesized, mechanically more
or less stable chainlike molecules or atomic structures
\cite{lehmann2002,visontai2010,emberly1999,khoo2008}, which are
contacted to some kind of leads at both ends, but otherwise
"free-hanging" without, e.g., connection to some substrate. There is
experimental data available on the conductivities of wires consisting of
a single molecule \cite{haiss2007,ashwell2006}. So far molecular wires
of lengths up to several nm could be experimentally produced.

Furthermore, there are various suggestions and realizations of
wire-like, 1-d,  nanostructures based on carbon, e.g., carbon nanotubes
or graphene-nano ribbons, placed on some substrate. Those carbon-systems
show remarkable electronic and transport properties (see, e.g.,
\cite{vandecasteele2009} and references therein). Both of them may
feature either metallic, semimetallic or semiconducting band structures
depending on chirality and width, respectively.

Some other experimental of quasi-$1$-d wires of nm-scale-diameter 
feature crystalline cores. Chemically stable silicon or metallic 
nanowires of that type have been
successfully fabricated \cite{ma2003,lin2004}. It has also been shown
by experiments \cite{ma2003} (scanning tunneling spectroscopy) and by
theory (density functional theory) \cite{rurali2005} that the electronic
structure of those nanowires is similar to the electronic structure of
the corresponding bulk material for diameters larger than $\propto 7$
nm. However electronic structure and resistivity may differ from
bulk-values for smaller diameters.

Yet another scheme envisions the formation of chains or rows of
conducting atoms or molecules on the surface of some insulating
substrate. The formation of such  linear molecular structures is thought
to be achieved by self-assembly, possibly influenced or controlled by
structures on the substrate surface. First steps in that direction have
been demonstrated \cite{kuhnle2004}.

Two points should be stressed here: i. So far none of the above 
techniques provides atomic wires which are scalable, stable and 
sufficiently controlled to set up complex electronic circuits. ii. 
Assume, nevertheless, arbitrarily long, stable wires were obtainable, 
then, according to all the above
suggestions, except for the first "free hanging" one, the electronic
system would be (weakly) coupled to bulk-type phonons. The latter being 
either
due to the lattice of the crystalline wire itself or simply to the
substrate. Even if impurities and electron-electron coupling were
negligible the phonon-coupling would inevitably control the transport
properties.

In the work at hand we essentially aim at analyzing the peculiarities of the
transport properties of some at present hypothetical, long atomic wires, 
the electrons of which are subject to scattering along the wire as 
mentioned under the above
second point.

Theoretically electronic transport through atomic wires is often
analyzed by using a Landauer-B\"uttiker-approach \cite{buettiker1988}.
Basically, the latter addresses a scenario of electrons moving
ballistically, i.e., unscattered, through an ideal $1$-d conductor from one
electrode to another (due to an applied voltage).
The resulting (finite) conductivity is mainly determined by the
properties of the contacts.
While this method is well suited for the above, relatively short, "free
hanging" wires the incorporation of rather complex
scattering along the wire appears challenging.

A standard approach to transport in bulk electron-phonon coupled systems 
is the Holstein approach. Here one relies on an evaluation of a
Kubo-formula by means of an approximate diagonalization of the system 
based on a transformation into a polaron picture. However, this
produces, in the limit of weak coupling relevant, diverging terms which 
would render the coupled model
ballistic. This can only be cured by the ad-hoc introduction of a
phenomenological (isotropic) polaron lifetime, the latter being due to
either impurity or electron-electron scattering, i.e., scattering
mechanisms which are not present in the pure electron-phonon model 
\cite{ortmann2009}.
Altogether this approach appears inadequate for the analysis of weakly
coupled but highly anisotropic electronic systems like the atomic wires
addressed below.

Our atomic wires may be viewed as hypothetical, ideal realizations of
the``self-assembled-atoms-type''. Our investigations are based on
substantially simplified models. We consider structures consisting of
several, parallel, infinite mono-atomic chains of atoms. Loosely
speaking the electrons may hop between the atomic sites, possibly with
different hopping strengths for the lateral and the longitudinal
direction. The band structure of such a system may be of the metallic,
insulating or semi-conducting type, (just like the graphene
nano-ribbons), depending on the number of rows and the hopping
strengths. We take the electrons as mutually non-interacting but weakly
coupled to some bulk phononic system. The sorrounding system is assumed
to be insulating, i.e., there is no electronic transport from the wire
into the sorrounding system. (While non-interacting electrons
are surely an idealization, this idealization may be adequate for 1-d
structures which tend to remain ballistic, even in the presence of
interactions \cite{zotos1999,narozhny1998,heidrich2003,zotos2003}).
Since, as explained above, neither the Landauer-B\"uttiker nor the
Holstein approach appear well-suited for such a system we resort to a
yet different approach based on Boltzmann equation descriptions.

This paper is organized as follows. In Sec.~\ref{sec-introduction} we
first introduce the above quantum model of the wire in detail. In
Sec.~\ref{sec-masterequation} we formulate, based on the model, a
linear(ized) master equation for the dynamics of some variables 
corresponding to occupation numbers of electronic momentum
modes. This is done on the basis of an approach derived in detail in 
\cite{bartsch2010-1}.
In Sec.~\ref{sec-diffusion} we suggest a Chapman-Enskog-type calculation 
of a
diffusion coefficient from a linear Boltzmann equation. By
means of the previously formulated master equation we set up such a
linear Boltzmann equation describing the electronic dynamics of our model.
Finally, in Sec.~\ref{sec-numerics} we numerically evaluate the obtained
formula for the diffusion coefficient of our model. We thus analyze the 
dependence of the
diffusion coefficient on the number of neighboring rows as well as on
the lateral coupling strength between neighboring rows, finding strongly 
discontinuous behavior.
We close with summary and outlook.

\section{Introduction of the Model} \label{sec-introduction}

Routinely we open the considerations by specifying the underlying quantum model on the basis of which we below analyze the dynamical properties of the atomic wires. The Hamiltonian consists of three parts: an electronic part $H_{0, \text{el}}$ describing the hypothetically unperturbed electrons on the wire, a phononic part $H_{0, \text{ph}}$ representing a standard set of decoupled phononic modes and an interaction $V$ which couples the electrons to the phonons:
\begin{eqnarray}
H\! &=& \!\underbrace{\sum\limits_{{\bf j}}\sum\limits_{r=1}^{B}\varepsilon_{{\bf j},r}a_{{\bf j},r}^{\dagger}a_{{\bf 
j},r}}_{H_{0, \text{el}}}\! +\!\underbrace{\sum_{{\bf i}}\omega_{{\bf i}}b_{{\bf i}}^{\dagger}b_{{\bf 
i}}}_{H_{0, \text{ph}}}\! \nonumber \\ 
&+& \!\underbrace{(\sum_{\substack{{\bf k,q}}}\sum_{\substack{r,s=1}}^{B}\frac{W_{rs}({\bf q})}{\sqrt{\Omega}}a_{{\bf k+q},r}^{\dagger}a_{{\bf k},s}b_{{\bf q}}\! +\! h.c.)}_{V}, 
\label{Hamiltonelph}
\end{eqnarray}
with $\Omega = BL$.

\begin{figure}[htb]
\hspace{-4.0cm}
\centering
\includegraphics[width=4.5cm]{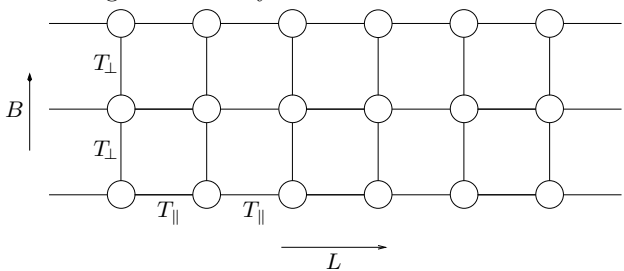}
\caption{Sketch of the quantum model we use to describe the properties of an atomic wire. Circles denote localized electronic states on a regular (quasi-)$1$-d lattice in configuration space. Lines indicate couplings with strength $T_{\|}$ parallel and strength $T_{\perp}$ perpendicular to the wire.} 
\label{fig0}
\end{figure}

A real space sketch of the electronic part is depicted in Fig.~\ref{fig0}. We assume a quasi-1-d, planar, cubic lattice of atoms that forms the wire, $L$ corresponds to the length. Concretely we assume that the electrons may occupy localized (Wannier) states at those lattice sites, i.e., a standard tight-binding description. The wire now consists of a finite, possibly small integer number $B$ of parallel rows of atoms. The motion of the electrons along the rows is controlled by nearest neighbor hopping matrix elements of strength $T_{\|}$. The hopping from one row to the adjacent row(s) is controlled by matrix elements of strengths $T_{\perp}$. The electrons on the wire are assumed to experience no mutual interactions, i.e., without any further coupling their dynamics would be simply ballistic. Of course, the tight-binding-description for the non-interacting electronic system is a strong idealization, which, however, may be a good approximation for 1-d systems.
 Since the wire is furthermore strictly periodic along the rows, its Hamiltonian may be denoted based on the dispersion relation $\varepsilon_{{\bf j},r}$ as done in (\ref{Hamiltonelph}). Here $\varepsilon_{{\bf j},r}
 $ denotes the one-particle-energy corresponding to the lattice-momentum $\bf j$ and the ``branch'' $r$. Due to the $B$ parallel rows the dispersion relation features $B$ branches or ``bands''.  Fourier transforming the Hamiltonian described by Fig.~\ref{fig0} essentially yields
\begin{equation}
\varepsilon_{{\bf j},r}= -2 T_{\|} \cos{\bf j}+T_{\perp}\varepsilon_r \ ,
\end{equation}
where the concrete $\varepsilon_r$ have to be determined from the subsequent diagonalization of a $B$x$B$-Matrix. (For an $B=2$ example see Fig.~\ref{fig1}.) Hence, the bandwidth of each band (which we label $E_B$) is determined by the longitudinal coupling $E_B = 4 T_{\|}$, whereas the displacement $T_{\perp}\varepsilon_r$ scales linearly with the lateral coupling $T_{\perp}$. That is, the model may feature a metallic, semiconducting or insulating band structure, depending on the choice of the system parameters, namely the hopping strengths $T_{\|}, T_{\perp}$. As it is demonstrated below, increasing $T_{\perp}$ for fixed $T_{\|}$ allows for a transition from metallic over semiconducting to insulating behavior.

The phononic dispersion relation $\omega_{{\bf i}}$ may take some standard form for acoustic or optical phonons. Below it will turn out that the details are irrelevant within the frame of our considerations as long as phononic energies are much smaller than electronic energies, i.e., $\omega_{{\bf i}} << \varepsilon_{{\bf j},r}$. The latter will be assumed.

The interaction $V$ is also of standard lowest order form, it may be viewed as corresponding to an electron scattering under creation or annihilation of a phonon, with the phonon essentially compensating for the electronic momentum change.
In our approach the electron-phonon-coupling $V$ is treated as a weak perturbation to the non-interacting electronic system.

\vspace{0.3cm}

\section{Formulation of a Master Equation for Coarse Grained Occupation Numbers} \label{sec-masterequation}
As a first step in our analysis we now aim at mapping the quantum dynamics of some set of ``coarse grained occupation number deviations'' in momentum space onto a master equation. To this end we follow directly the scheme described in \cite{bartsch2010-1}. For an overview over mappings of quantum dynamics onto Boltzmann equations also see \cite{bartsch2010-1} and references therein. This scheme produces a linear master equation for a finite number of variables. These variables are sums of deviations of electronic occupation numbers from their equilibrium values. To make this well defined we have to specify which occupation number deviations should contribute to a certain sum. Only with this specification the scheme from \cite{bartsch2010-1} may be implemented. We call all modes, the occupation number deviations of which contribute to a certain sum, a ``grain''.  Thus the number of variables as well as the concrete rates from the above master equation depend on this graining. Since this graining is to some extent arbitrary, we must make sure that eventually our diffusion coefficient converges for a systematic refinement of the graining. This convergence then in turn sets the scale for a ``correct'' graining. In the following we first describe our choice for the graining and then give precise definitions for the above dynamical variables of the master equation.

\begin{figure}[htb]
\centering
\includegraphics[width=9cm]{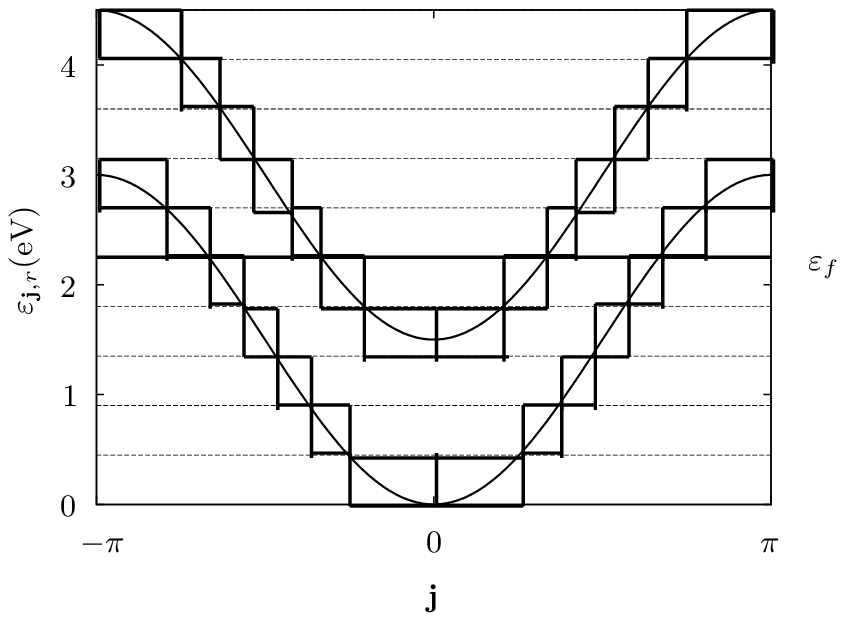}
\caption{Sketch of our specific coarse graining in momentum space for a model featuring $B=2$ rows, i.e., the dispersion relation consists of $2$ bands. Parallel dashed lines define energy shells of equal energy width $\Delta E$. Rectangular boxes mark the individual grains. The Fermi edge is indicated by a thick line. Some shells contain grains from one band, some from both bands.
Other model parameters: $E_B = 3 \text{eV}$, $T_{\perp}=T_{\|}$, $M=10$ energy shells and half filling ($\varepsilon_{f} = 2.25 \text{eV}$). 
}
\label{fig1} 
\end{figure}

Our graining is realized by primarily partitioning the momentum space into $M$ ``energy shells'' of equal energy width $\Delta E$. We then define the grains within one shell by means of the intersections of the electronic dispersion relation with the respective energy shell as depicted exemplarily in Fig.~\ref{fig1}.
We label the entity of all grains serially by Greek indices (e.g., $\kappa$).
(This may correspond to a consecutive numbering from left to right (see Fig.~\ref{fig1}) within one shell, which is then accordingly continued through the shells with increasing energy.)
Note that these grain indices already contain the information to which band the respective grain belongs, that is, the band indices $r$ are absorbed into the grain indices. As mentioned above the necessary refinement of the graining is eventually determined by the convergence of the diffusion coefficient. However, we generally focus on grains that are small enough to allow for a linearization of the dispersion relation on each grain.

\vspace{0.3cm}

Now we give a mathematical definition for the above deviations of electronic occupation numbers from their equilibrium values which are summed over a grain to form a dynamic variable of the above master equation. An operator representing such a deviation is given by 
\begin{eqnarray}
\Delta_{{\bf j},r}:=
a_{{\bf j},r}^{\dagger}a_{{\bf j},r}-f_{{\bf j},r}, 
\end{eqnarray}
where $f_{{\bf j},r}$ is the equilibrium value of the electronic occupation number. Since we are interested in weak electron-phonon coupling, we may take $f_{{\bf j},r}$ from the Fermi distribution for the occupation numbers of the isolated electronic system. Since we are furthermore interested in low temperature 
regimes, we may routinely replace the chemical potential $\mu$ in the corresponding Fermi distributions $f_{{\bf j},r}(\mu,T)=(\exp ((\varepsilon_{{\bf j},r}-\mu)/k_{B}T)+1)^{-1}$ by the Fermi energy $\varepsilon_{F}$. With the help of such ``individual mode deviation operators'' $\Delta_{{\bf j},r}$ the deviation operators for the whole grains $\Delta^{\kappa}$ may simply be defined by the corresponding sums
\begin{equation}
\Delta^{\kappa}=\sum_{{\bf j}, r\in\kappa} \Delta_{{\bf j},r} \ , 
\end{equation}
where the sum of course runs over all modes belonging to some grain. The dynamical variables of the above master equation are then defined as the expectation values of the grain deviation operators $d^{\kappa}(t)$ for the actual quantum state $\rho (t)$ of the full system evolving according to the coupled Hamiltonian $H$, i.e., $d^{\kappa}(t):=\text{Tr}\{ \rho (t)\Delta^{\kappa}\}$. Naturally, the dynamics of $\rho (t)$ are not directly accessible. But, as explained above, the considerations in \cite{bartsch2010-1} (which we, for brevity, do not intend to repeat or discuss here) provide a scheme by means of which the dynamics of the $d^{\kappa}(t)$ may be mapped onto a linear master equation. The scheme is believed to be valid as long as the electron-phonon coupling may be considered a small perturbation to the electronic system $H_{0, \text{el}}$. Thus, simply applying this scheme casts the dynamics into the form: 
\begin{equation}
\dot{d}^{\eta}(t)= \sum_{\kappa\neq\eta} R_{\eta\kappa}(t) d^{\kappa}(t)-\sum_{\kappa\neq\eta} R_{\kappa\eta}(t) d^{\eta}(t) \ .
\end{equation}
The concrete expressions of the rates may also be read off from the results of \cite{bartsch2010-1}. As mentioned above we focus here on the case of negligible phonon energies which is explicitly treated in \cite{bartsch2010-1}. Exploiting this leaves us with
\begin{eqnarray}
\label{rateimp}
&&R_{\eta\kappa}(t)= \nonumber\\
&&\hspace{-0.5cm}\int_{0}^{t}\!\!\!\text{d}\tau \frac{2}{\hbar^{2}N_{\kappa}}\sum_{\substack{{\bf i}, s\in\kappa,\\{\bf k}, r\in\eta}}\frac{\vert W_{rs}({\bf k}-{\bf i})\vert ^{2}}{\Omega}(1\! +\! 2 g_{{\bf k}-{\bf i}}\!) \nonumber\\
&&\cdot\cos [\frac{1}{\hbar}(\varepsilon_{{\bf i}, s}\! -\!\varepsilon_{{\bf k}, r})\tau], \nonumber\\
\end{eqnarray}
where $N_{\kappa}$ denotes the number of states within the grain $\kappa$ and $g_{{\bf q}}$ is the equilibrium occupation number of the phonon mode ${\bf q}$, as taken from the Bose distribution for the isolated phononic system $H_{0, \text{ph}}$. (Note that the phononic occupation numbers are not treated as dynamical variables in our approach.)
Basically, the rates appear as integrals over some correlation functions, which usually decay within some correlation time $\tau_c$, we generally assume the latter to be short compared to the time scale of the relaxation dynamics (weak coupling limit). 
Further, we define
\begin{equation}
\overline{\Gamma_{\eta\kappa}^{2}}=\sum_{\substack{{\bf i}, s\in\kappa,\\{\bf k}, r\in\eta}}\vert W_{rs}({\bf k}-{\bf i})\vert ^{2}(1\! +\! 2 g_{{\bf k}-{\bf i}}\!)\cdot\frac{1}{N_{\kappa}N_{\eta}}\ .
\end{equation} 
Using this abbreviation and routinely exploiting the properties of the $sinc$-function as well as the linearizability of the dispersion relation on each grain, we can approximately perform the integration in (\ref{rateimp}), finding for times larger than the correlation time for the rates 
\begin{equation}
\label{ratma}
R_{\eta\kappa}=\frac{2\pi \overline{\Gamma_{\eta\kappa}^{2}}}{N \hbar}\frac{N_{\eta}}{\Delta E}\delta_{E(\eta),E(\kappa)}\ .
\end{equation}
The $\delta_{E(\eta),E(\kappa)}$ is meant to indicate that transitions only occur between grains from the same energy shell, i.e., all other rates vanish. Note that the resulting rates are completely independent of the electronic equilibrium occupation numbers, i.e., scattering above, below, or at the Fermi-level is in this approximation qualitatively the same.

\section{Calculation of a diffusion coefficient from a linear Boltzmann equation} \label{sec-diffusion}

Eventually we intend to calculate the diffusion coefficient from a type of linear Boltzmann equation. To this end we are simply going to interpret the rates from (\ref{ratma}) as the entries of the corresponding collision term. But even if the Boltzmann equation is fully defined, we have to specify how to calculate a diffusion coefficient from it. There are several approaches to the derivation of diffusion coefficients from linear(ized) Boltzmann equations \cite{balescu1975,brenig1989,jaeckle1978,kadiroglu2007}. However, since we are here confronted with a subtlety that may not occur in other cases, namely a non-symmetric rate matrix, we suggest an alternative approach to the diffusion coefficient. This approach is outlined in the following.

We start from a linear Boltzmann(-like) equation describing the dynamics of some probability distribution $\rho_{\eta}({\bf x}, t)$ in a velocity discretized one-particle phase space: 
\begin{equation}
\dot{\rho}_{\eta}({\bf x}, t)=-v_{\eta}\nabla_{\bf x}\rho_{\eta}({\bf x}, t)+\sum_{\kappa}R_{\eta\kappa}\rho_{\kappa}({\bf x}, t)\ .
\end{equation}
The two addends represent the drift and the collision term, respectively. The velocity $v$ appears as depending on some index $\eta$. Eventually this index will be identified with the quantum mechanical momentum mode.
After Fourier transforming with respect to  ${\bf x}$ we find for the (Fourier transformed) probability distribution $\rho_{\eta}:=\rho_{\eta}({\bf q}, t)$
\begin{equation}
\dot{\rho}_{\eta}=\sum_{\kappa}(\underbrace{-\imath {\bf q} v_{\eta} \delta_{\eta\kappa}}_{R_1}+ \underbrace{R_{\eta\kappa}}_{R})\rho_{\kappa}\ ,
\label{bolschr}
\end{equation} 
where the drift term $R_1$ now enters as a diagonal matrix and the collision term $R$ as a non-diagonal matrix with respect to the above representation.

Eq.~(\ref{bolschr}) is formally identical to a non-specific, abstract time dependent Schr\"odinger equation where $R_1 +R$ corresponds to some Hamilton operator. However, since $R_1 +R$ is not Hermitian, eigenvalues may (and will) not be real. Furthermore, eigenstates represent spatial particle-density waves with wavevectors $\bf q$ which decay mono-exponentially. Such a behavior is in accord with diffusive transport. The corresponding decay constant is given by the imaginary part of the corresponding eigenvalue. For arbitrarily small ${\bf q}$ (long wavelengths) one may treat the drift term $R_1$ as a small perturbation to the collision term $R$.
Therefore, in order to find the decay constant, we may calculate corrections to eigenvalues of $R$ induced by $R_1$ via a procedure that is essentially analogous to standard time-independent perturbation theory as given in textbooks. In doing so, the eigenvectors of $R$ represent the eigenstates of the unperturbed Hamiltonian. The eigenvector $\vec{P}_0$ of $R$ featuring the eigenvalue zero corresponds to the total, isotropic equilibrium distribution in the above phase space. Thus, assuming that the evolving distribution that corresponds to a diffusion process exhibits  locally (with respect to real space) more or less the equilibrium distribution with respect to velocities (momenta), we have to analyze the decay dynamics of $\vec{P}_0$ induced by $R_1$. In other words we have to calculate the correction to the eigenvalue of $\vec{P}_0$. Before going in any detail we note that the second order correction scales as $\propto {\bf q}^2$. Thus, in case the first order vanishes, the
  decay constants for the corresponding density waves will scale as $\propto {\bf q}^2$ as well. The latter dynamics represent a solution to a diffusion equation with a diffusion coefficient given by the prefactor of the ${\bf q}^2$-term of the decay constants. Thus, calculating the second order correction to $\vec{P}_0$ may and will turn out as being equivalent to calculating the diffusion coefficient. 

Nevertheless, as already mentioned above, we encounter here the difficulty that the collision term $R$ is in general non-symmetric and therefore the eigenvectors of $R$, denoted by $\{ \vec{P}_{i}\}$, are not necessarily orthonormal. This problem may be dealt with by introducing a  dual basis $\{ \vec{P}^{i}\}$ such that 
\begin{equation}
\vec{P}^{j}\vec{P}_{i}=\delta_{j i}\ . 
\end{equation}
Like, e.g., in the context of relativity, the dual basis is constructed by means of some metric tensor $G_{i j}$ featuring
\begin{equation}
\vec{P}^{i}=G_{i j}\vec{P}_{j}\ .
\end{equation}
A straightforward calculation yields for the inverse of the metric tensor 
\begin{equation}
G^{-1}_{i j}=\vec{P}_{i}\cdot\vec{P}_{j}\ .
\end{equation}
That is, its matrix elements are simply given by the scalar products of the eigenvectors of $R$.
For the construction of the dual basis $G^{-1}$ has still to be inverted of course.

Within this framework we are finally able to determine the corrections to the eigenvalue zero. As already outlined, we hereby follow precisely the scheme of standard time independent perturbation theory, except for replacing basis ``bra''-vectors by vectors from the dual basis. For the eigenvalue correction (to the eigenvalue of $\vec{P}_0$, i.e., zero) to first order we thus obtain
\begin{equation}
E_1 = -i {\bf q} \vec{P}^0 \hat{v} \vec{P}_0 \ ,
\end{equation}
where $\hat{v}$ is a diagonal matrix in the matrix representation used in (\ref{bolschr}) with the corresponding velocities $v_{\eta}$ as diagonal elements, i.e., it is given by $R_{1}=-\imath{\bf q} \hat{v}$.
If the system in question features full spatial mirror-symmetry (as our atomic wire system does) the eigenvectors $\vec{P}_{j}$ of  $R$ as well as their dual counterparts $\vec{P}^{i}$ separate into two symmetry classes, symmetric and antisymmetric. Those two classes do not mix or overlap (in the sense of a regular scalar product). Since $\vec{P}_0$ and $\vec{P}^0$ belong to the symmetric class but $\hat{v} \vec{P}_0$ belongs to the antisymmetric class, the first order correction $E_1$ vanishes. As outlined above this is exactly what is required for the corresponding solution of the Boltzmann equation in order to describe diffusive dynamics at all. Thus we may indeed extract the diffusion coefficient from the second order. A straightforward calculation yields for the latter 
\begin{equation}
D =\frac{E_2}{-{\bf q}^2} =-P^{0}_{\eta}v_{\eta}R^{-1}_{\eta\kappa}v_{\kappa}P_{0,\kappa} \ ,
\label{difco}
\end{equation}
where $R^{-1}$ corresponds to the inverse matrix to $R$ neglecting the null-space.

Thus we eventually found a diffusion coefficient describing a diffusive class of solutions of a velocity-discretized, linear Boltzmann equation. Or to rephrase: given a specific linear Boltzmann equation we may calculate the corresponding diffusion coefficient based on (\ref{difco}). Thus, in our case we now have to specify a pertinent Boltzmann equation, i.e., the $ v_{\eta}$ and $R_{\eta\kappa}$. Of course we do that on the basis of the considerations in Sec.~\ref{sec-masterequation}.  We formulate a Boltzmann  equation which is discretized according to the previously described graining. Concretely we identify the $ v_{\eta}$ with the slope of the dispersion relations at the centers of the corresponding grains. This is the standard concept of identifying particle velocities with group velocities. The scattering rates $R_{\eta\kappa}$ are taken from (\ref{ratma}). With those allocations the Boltzmann equation is fully specified. However, since the rates as given in (\ref{ratma}) only give rise to transitions within an energy shell, the dynamics completely decouple with respect to energy shells. Due to this decoupling the equilibrium state $\vec{P}_{0}$ is not completely determined by the Boltzmann equation, the latter does not fix the equilibrium particle numbers for full shells, it only sets the distribution onto the grains within one shell. Thus we have to pick a $\vec{P}_{0}$ in accord with the Boltzmann equation but additionally with reasonable overall particle numbers assigned to the shells. Here we do that based on the standard idea of transport being due to particles above the Fermi edge and due to holes below. Thus we weigh the equilibrium distribution within the shells by the probabilities of finding holes or particles, respectively at the corresponding energies. Doing so we find:
\begin{equation}
\vec{P}_0: P_{0,\kappa}\propto f(E,T)N_{\kappa}\ 
\end{equation}
for $E>\varepsilon_{f}$ and
\begin{equation}
\vec{P}_0: P_{0,\kappa} \propto (1-f(E,T))N_{\kappa}
\end{equation}
for $E<\varepsilon_{f}$.
This construction ensures that the diffusion coefficient is dominated by processes occurring near the Fermi energy for common temperatures which appears physically reasonable. Since the diffusion coefficient as given in (\ref{difco}) involves the dual counterpart of the equilibrium distribution, an overall prefactor, i.e., the total number of particles does not alter its value (, hence such a prefactor is not specified above). This also appears physically reasonable. With this choice for the equilibrium distribution all quantities in (\ref{difco}) are eventually specified.

\section{Numerical computation of the diffusion coefficient} \label{sec-numerics}

In this paragraph we concretely determine a diffusion coefficient for the model describing the atomic wire as introduced in Sec.~\ref{sec-introduction} by explicitly and straightforwardly evaluating formula (\ref{difco}).
The respective computations may be done on standard computers. The numerical effort of course depends on the finesse of the chosen/necessary graining, however most of the diffusion coefficients presented below have been found within minutes of computation time.

The calculations mainly aim at describing the typical, characteristic dependence of the diffusion coefficient on two crucial system properties, i.e., the width of the atomic wire and the lateral coupling; we do not intend here 
to determine diffusion coefficients with extreme (numerical) precision or to describe concrete experimental realizations in great detail. 

Along those lines the bandwidth of the electronic dispersion relation is chosen as $E_B=3\text{eV}$, i.e., $T_{\|}=0.75\text{eV}$.  
Furthermore, we apply a temperature of $300\text{K}$. In this case the equilibrium distribution $\vec{P}_0$ is mainly concentrated around the Fermi energy, i.e., mainly the grains belonging to the shells in the vicinity of the Fermi edge contribute to the diffusion coefficient. Furthermore, we exclusively focus here on half-filling situations. For simplicity we assume that the electron-phonon coupling elements $W_{rs}({\bf q})$ from (\ref{Hamiltonelph}) show no strong ${\bf q}$-dependence and also that there is no decisive dependence on the band indices $r,s$, i.e., we consider all momentum modes as being equally strongly coupled, no matter which grain they belong to. Consequently, we set $W_{rs}({\bf q})^2 = W^2$ with $W=0.03\text{eV}$, thus weak coupling is implemented in the sense of electron-phonon coupling being much smaller than the electronic bandwidth. Additionally, also for simplicity, we set the phononic dispersion to a constant with $\omega_{\bf i}=0.03\text{eV}$. 
 This satisfies the above mentioned requirement of phonon energies being much smaller than electronic energies.

As outlined above, we expect the diffusion coefficient to converge towards some value for finer and finer grainings. We clearly observe this feature in the calculations. 
So, all plotted diffusion coefficients refer to in this sense converged values, i.e., are taken from sufficiently fine grainings which here means about $100$ shells per bandwidth $E_B$. This amounts to a total number of grains $g\approx 200 B$.

Fig.~\ref{fig3} shows the dependence of the diffusion coefficient on the lateral coupling $T_{\perp}$ for several numbers of rows ($2$ to $6$).

\begin{figure}[htb]
\centering
\hspace{-1.0cm}
\subfigure[]{\label{fig-disp2}
\includegraphics[width=7.0cm]{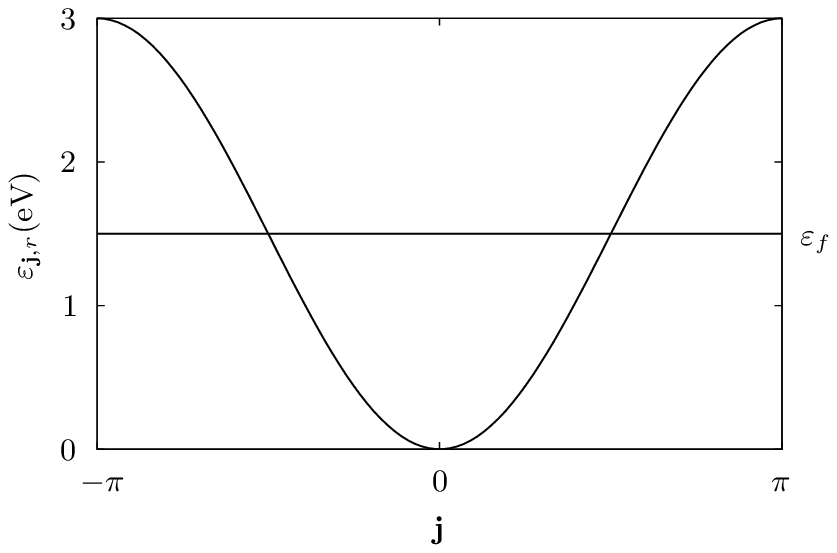}}
\hspace*{-1.0cm}
\subfigure[]{\label{fig-disp3}
\includegraphics[width=7.0cm]{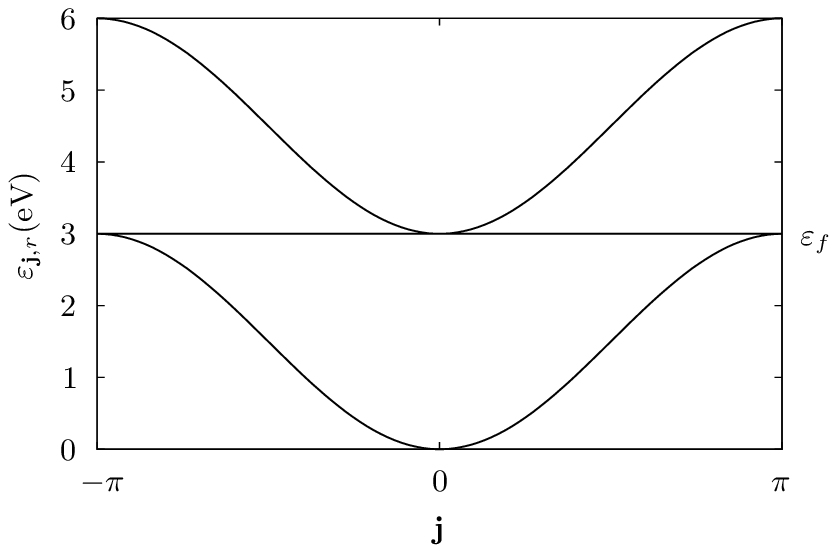}}
\subfigure[]{\label{fig-disp4}
\includegraphics[width=7.0cm]{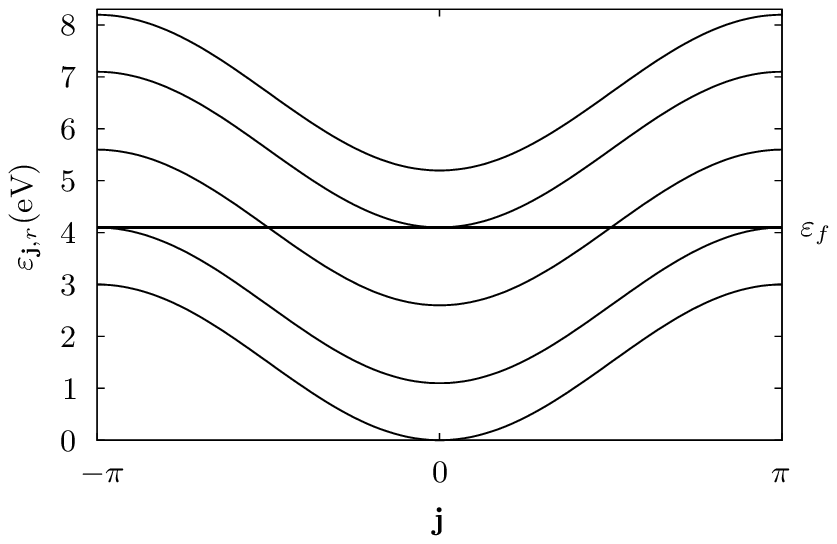}}
\hspace{0.8cm}
\caption{Dispersion relation for (a) $1$ row, $T_{\|}=0.75\text{eV}$, (b) $2$ rows, $T_{\|}=0.75\text{eV}$, $T_{\perp}=2 T_{\|}$, (c) $5$ rows, $T_{\|}=0.75\text{eV}$, $T_{\perp}=2 T_{\|}$.} 
\label{graphs}
\end{figure}

The drops of the diffusion coefficient in Fig.~\ref{fig3} with increasing lateral coupling may be interpreted in terms of a metal-semiconductor-insulator transition with respect to the characteristics of the underlying dispersion relation.
Fig.~\ref{fig-disp2} shows the dispersion relation for one row and, simultaneously, for any number of rows with $T_{\perp}=0$. (In the latter case the band structure consists of a number of branches equal to the number of rows, which all coincide.) So Fig.~\ref{fig-disp2} refers to the utmost left data points in both Figs.~\ref{fig3} and \ref{fig2}.
Here, the band structure is clearly metallic.
For $2$ rows the corresponding band structure is still metallic for $T_{\perp}= T_{\|}$ (see Fig.~\ref{fig1}), where the Fermi edge crosses both branches, while there is a transition to semiconducting behavior for $T_{\perp}=2 T_{\|}$ (see Fig.~\ref{fig-disp3}), since the Fermi energy now lies between the two bands. 
(Recall that $T_{\perp}$ determines the displacement of the different energy bands.)
This is in accord with the decrease of the diffusion coefficient to very small values for
$T_{\perp}=2 T_{\|}$. Further increasing of $T_{\perp}$ would lead to the opening of a band gap and eventually to an insulating band structure.

As it may be observed in Fig.~\ref{fig3} the diffusion coefficients generally decrease to small values for all depicted widths, which is then followed by jumps up to higher values at certain lateral coupling strengths, the latter varying significantly for different widths.
The jumps occur whenever a band edge is driven through the Fermi edge, since the slope of the dispersion relation is $0$ at the band edges, which leads to very large rates $R_{\eta\kappa}$. This feature may be illustrated by the example of $5$ rows with $T_{\perp}=2 T_{\|}$ (see Fig.~\ref{fig-disp4}), which is also covered by the data of Figs.~\ref{fig3} and \ref{fig2}. 
Although the band structure is in principle metallic because the Fermi edge crosses one band, the diffusion coefficient is still very small, since the Fermi edge lies on the edge of two other bands, which yields high rates $R_{\eta\kappa}$. If 
$T_{\perp}$ is now further increased, those two bands are soon driven out of the direct vicinity of the Fermi edge and the corresponding scattering channels (i.e., states into which scattering processes may occur,) get out of reach for not too large temperatures. This is what leads to a ``sudden'' increase of the diffusion coefficient.
Especially note that these jumps cause a significant spreading of the curves above $T_{\perp} \approx 1.3 T_{\|}$, i.e., when the lateral coupling becomes stronger than the longitudinal. 

This feature is further analyzed in Fig.~\ref{fig2}, which shows the resulting diffusion coefficients for different widths of the atomic wire, i.e., different number of parallel rows $B$, for two lateral couplings $T_{\perp}=2 T_{\|}$ and $T_{\perp}=T_{\|}$. 
One finds that the corresponding diffusion coefficients for $T_{\perp}=2 T_{\|}$  come out rather discontinuous 
and become very small for certain widths, which may be explained analogously to the examples discussed above ($T_{\perp}=2 T_{\|}$, $2$ and $5$ rows). This effect diminishes for larger numbers of rows, since there are many bands which are crossed by the Fermi edge and only few which are tangent to it.

For $T_{\perp}=T_{\|}$ one finds a relatively smooth, continuous curve. 
However, in both cases we observe a dependency of the diffusion coefficient on the width of the atomic wire, which has to be classified as a finite size effect. Of course this effect should vanish in the limit of infinitely large widths $B\rightarrow\infty$, i.e., the diffusion coefficient is expected to become constant, since the model then passes into a regular $2$-d lattice.
The plots in Fig.~\ref{fig2} suggest such a convergence at a width of approximately $50$ rows and are therefore in accord with the above educated guess. Although a numerical verification was possible we omit such a calculation here for it would be numerically rather costly.

The above argument concerning the influence of the properties of the electronic band structure on the investigated dependencies of the diffusion coefficient is also valid for models with more realistic dispersion relations or electron-phonon-couplings, so that the discontinuous behavior will most likely remain. The same holds true for scattering mechanisms other than electron-phonon, such as weak impurity or electron-electron scattering.

\begin{figure}[htb]
\centering
\includegraphics[width=7cm]{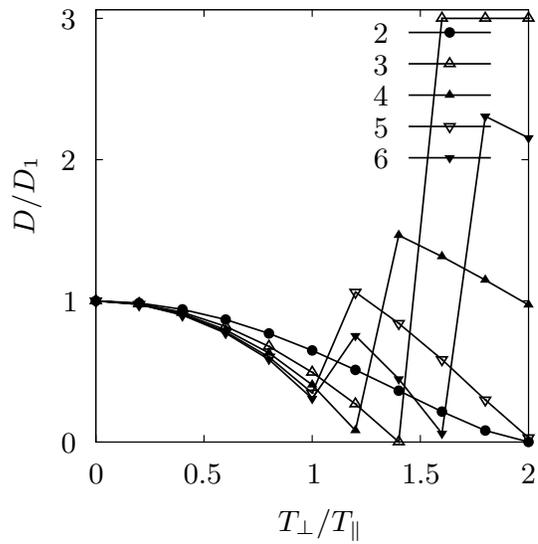}
\caption{Plot of the diffusion coefficient $D$ against the lateral coupling $T_{\perp}$ for different widths, i.e., different numbers of rows, according to the legend. The plot is normalized to the $D$ obtained for $1$ row, denoted by $D_1$. Other parameters: $E_B = 3\text{eV}$, $T_{\|}=0.75\text{eV}$, $T=300\text{K}$, half filling.
} 
\label{fig3}
\end{figure}

\begin{figure}[htb]
\centering
\includegraphics[width=7cm]{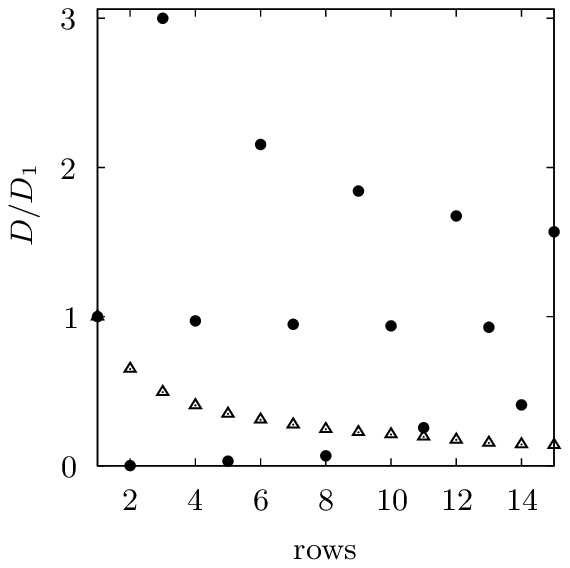}
\caption{Plot of the diffusion coefficient $D$ against the number of rows, i.e., the width for $T_{\perp}=2 T_{\|}$ (black dots) and $T_{\perp}= T_{\|}$ (white triangles). The plot is normalized to the $D$ obtained for $1$ row, denoted by $D_1$. Other parameters: $E_B = 3\text{eV}$, $T_{\|}=0.75\text{eV}$, $T=300\text{K}$, half filling.
}
\label{fig2}
\end{figure}

\section{Summary and Outlook} \label{sec-summary}

We investigated diffusive transport behavior of electrons in atomic wires, described on the basis of some pertinent quantum model. The rather simplified model is meant to schematically describe planar atomic wires with a width of a few atoms, formed, e.g., on a crystal surface (or embedded in some crystal bulk). The electrons within the wire are taken to be  mutually non-interacting, diffusive behavior is solely induced by coupling to the crystal phonons. 
To this end we firstly computed a linear master equation describing the dynamics for some, with respect to momentum space, coarse grained occupation numbers, hereby closely following an approach from \cite{bartsch2010-1}.
The resulting rate matrix is then interpreted as the collision term of a pertinent linear Boltzmann equation. 
Furthermore, we analyzed solutions of linear Boltzmann equations which can be characterized as representing diffusive transport dynamics. An explicit expression connecting the diffusion coefficient to the parameters of the Boltzmann equation is given. Based on this formula we numerically computed diffusion coefficients for wires of different types. 
In particular, we investigated the dependency of the diffusion coefficient on the width of the atomic wire
as well as the dependency on the lateral coupling strength between neighboring rows.
We found that,
if the lateral exceeds the longitudinal atomic coupling within the wire, the diffusion coefficient exhibits large jumps with respect to varying widths.
This feature may be clearly classified as a finite size effect. Concretely the diffusion coefficient may become very small and even nearly vanish for certain combinations of widths and lateral couplings.
There is also a rather strong dependence on the lateral coupling for fixed widths showing jumps for strong lateral coupling. The corresponding curves differ significantly subject to the width. Thus, to summarize this results: the diffusion coefficient depends strongly and in a non-trivial way on both, the wire width and the lateral coupling.
We additionally discussed how this discontinuous dependence may be understood in terms of the band structure of the non-interacting electronic system.

By using a generalized Einstein relation \cite{steinigeweg2009} the diffusion coefficients may be converted into conductivities, which would probably be better accessible by experiments. Therefore, the approach introduced here may serve as a basis for concrete transport-theoretical analyzations to be performed in the near future. 

To obtain more realistic quantitative results, the model parameters used in this approach may be better adjusted to some concrete (experimental)
realizations, e.g., a more realistic electron-phonon coupling, possibly including electron-electron interactions, impurities etc..
Additionally, one could choose a more realistic electronic dispersion relation, which may, e.g., be obtained by density functional theory. 

We close with the following rather speculative suggestion: One may also think of possible applications of such atomic wires as controlled switches or transistors on the nanoscale. If such an atomic wire consisting only of a few atomic rows is operated at a point where the Fermi edge and the band edge of one electronic branch coincide (as depicted, e.g., in Fig.~\ref{fig-disp3}), there is obviously a very strong dependence of the transport coefficient on small variations of the band structure, see, e.g., the case of $B=3$, $T_{\perp}/ T_{\|} = 1.3$ in Fig.~\ref{fig3}. Such a variation of the band structure may be induced in a controlled fashion. This could be achieved by a local lateral deformation of the wire through some piezoelectric mechanism which causes a variation of the lateral coupling. More promising may be the concept of applying a lateral potential gradient by means of some gate electrode. Such a potential gradient would also have a direct effect on the relative positions of the electronic branches in the energy scheme and hence induce a strong change of the transport coefficient. Note that such a transistor could be implemented without any form of doping and, just like a field effect transistor, would not require any gate current.

%
%

\begin{acknowledgments}

We sincerely thank M. Kadiro\=glu for his contributions to fruitful
discussions. Financial support by the ``Deutsche
Forschungsgemeinschaft'' is gratefully acknowledged.

\end{acknowledgments}

%
%

\end{document}